\documentclass[prb,twocolumn,showpacs,preprintnumbers,amsmath,amssymb]{revtex4}
\usepackage{graphicx}% Include figure files
\usepackage{dcolumn}% Align table columns on decimal point
\usepackage{bm}% bold math

\newcommand{\rv}{{\bf r}}

\newcommand{\pv}{{\bf p}}

\begin{document}

\title{Monte Carlo transient phonons transport in silicon and germanium at nanoscales}

\date{25 March 2005}

\author{David Lacroix}
\email{David.Lacroix@lemta.uhp-nancy.fr}
\altaffiliation[Also at ]{Laboratoire d'\'Energ\'etique et de M\'ecanique Th\'eorique et Appliqu\'ee\\ 
Universit\'e Henri Poincar\'e, Nancy 1\\ 54506 Vand\oe{uvre} Cedex, France}

\author{Karl Joulain}
\email{Karl.Joulain@ensma.fr}
\affiliation{Laboratoire d'\'Etudes Thermiques\\ ENSMA\\ 1, Avenue Cl\'ement Ader
86960 Futuroscope Cedex, France}

\begin{abstract}
Heat transport at nanoscales in semiconductors is investigated with a statistical method. The Boltzmann Transport Equation (BTE) which characterize phonons motion and interaction within the crystal lattice has been simulated with a Monte Carlo technique. Our model takes into account media frequency properties through the dispersion curves for longitudinal and transverse acoustic branches. The BTE collisional term involving phonons scattering processes is simulated with the Relaxation Times Approximation theory. A new distribution function accounting for the collisional processes has been developed in order to respect energy conservation during phonons scattering events. This non deterministic approach provides satisfactory results in what concerns phonons transport in both ballistic and diffusion regimes. The simulation code has been tested with silicon and germanium thin films; temperature propagation within samples is presented and compared to analytical solutions (in the diffusion regime). The two materials bulk thermal conductivity is retrieved for temperature ranging between 100 K and 500 K. Heat transfer within a plane wall with a large thermal gradient (250 K-500 K) is proposed in order to expose the model ability to simulate conductivity thermal dependence on heat exchange at nanoscales. Finally, size effects and validity of heat conduction law are investigated for several slab thicknesses.
\end{abstract}

\pacs{05.10.Ln,\,44.10.+i,\,63.20.-e,\,65.40.-b}

\maketitle

\section{Introduction}
The development of nanotechnologies has lead to an unprecedented size reduction
of the electronic and mechanical devices. For example, transistors of a few nanometer size
are now openly  considered \cite{itrs04}. The heat that will be dissipated by joule effect in these semi-conductors junctions will reach soon the levels of the heat dissipated in a light bulb.  This high volumetric heat dissipation in electronic devices will have to be evacuated very efficiently in order to avoid possible failures of the systems. This task will not be achieve without a sharp knowledge of the phenomena governing the heat transfer at nanoscale. Furthermore, new technologies based on a local heating are  being developed in order to enlarge the computer hard disk capacity. The ultimate limit of storage is to write a byte at the atomic scale. This goal is already feasible with near-field microscope probes but at a too slow rate. A way to  write bytes at the nanometer scale is the melting of a polymer by heating it on a very short time scale ($\le 1$ ns) by an array of heated  near-field probes \cite{Vettiger02}.  In this example, the heat transfer has to be controlled not only at the nanometer scale but also at the nanosecond scale. 

Through these two examples, one can anticipate that the foreseeing technological challenges in miniaturization will have to solve more and more problems of heat transfer at short time and space scale.
However, the physics of heat transfer usually used (Fourier's law, Radiative Transfer Equation) can no longer be applied when some characteristic length scales are reached \cite{Cahill03}. In thermal radiation for example, wave effects appear as the system characteristic lengths becomes lower than the typical wavelength ($\lambda\sim 10\mu$m at $T=300$ K) \cite{Polder71,Loomis94}. A substantial increase of the radiative heat transfer can even be reach at nanometric distances \cite{Mulet01}. On its side, heat conduction is classically described by the Fourier law and the heat conduction equation ($\partial T/\partial t=\alpha\Delta T$) which is a diffusion equation. It is well known that this kind of equation can be interpreted as a random walk of particles \cite{Einstein05}. In the case of heat conduction, we are actually dealing with energy carriers which are electrons in metals and phonons in crystalline materials. When these carriers undergo a large number of collisions, the use of the diffusion equation is valid whereas a more careful study is required when the number of interactions between carriers lowers.

A way to achieve this goal is to consider the evolution of a distribution function $f(\rv,\pv,t)$ which describes the number of particles in a certain elementary volume $d^3\rv d^3\pv$ around the point ($\rv,\pv$) in the phase space. The evolution equation of $f$, called the Boltzmann Transport Equation (BTE) makes $f$ to vary in space and time under the influence of advection, external force and collision \cite{Ziman}. Note that this approach is not relevant to treat the wave aspects of the problem such as interference or tunneling. The understanding and the modeling of the collision term is actually the key point in the resolution of the BTE. It can sometimes be fully modelized as in radiation transfer. Then the collision term is in that case the sum of an absorption term, an elastic scattering term and an emission term proportional to an equilibrium distribution \cite{Chandrasekhar}. Many resolution technique have been developed in radiation transfer such as the Discrete Ordinates Method, the  Monte Carlo method or the ray-tracing method \cite{Modest}. They can hardly be used when the collision processes are inelastic as it is the case for electrons and phonons \cite{Landau_kin}. For example, the phonons, which are eigenmodes of the harmonic oscillators constituting the crystal, can only interact through the anharmonic term of the potential leading to three or more phonon collisions. These interactions preserve neither the number of phonons nor their frequency in the collision process. Nevertheless, these three or four phonons interactions tend to restore thermal equilibrium i.e. to help the phonons to follow an equilibrium distribution function which can be easily determined from thermodynamic equilibrium considerations. Thus, much of the studies modelize the collision term in the BTE by the so-called Relaxation Time Approximation : the distribution function $f(\rv,\pv,t)$ relaxes to an equilibrium function $f^0(\rv,\pv)$ on a time scale $\tau(\pv)$. The BTE resulting from this approximation is nothing but the Radiative Transfer Equation without scattering\cite{Majumdar93}. All the numerical tools developed in thermal radiation can therefore be used in this case. The key point in this model is to calculate a suitable $\tau(\pv)$ in order to characterize the collisions. 

In the middle of the twentieth century, a great theoretical effort has been done to modelized the relaxation times of the phonons in a  bulk material. At ambient temperature, it has been shown that the main contribution to the relaxation time finds its origin in the anharmonic phonon interaction. Two kinds of processes can be identified. The so-called Normal processes ($N$) which maintain the momentum in the collision and the Umklapp processes ($U$) which do not preserve the momentum. The former do not affect the material thermal resistance contrary to the latter. These Umklapp processes  follow selection rules \cite{Klemens51} and it is an amazing feat to calculate them \cite{Herring54}.  Actually, the relaxation times are known for a small number of materials and often in the bulk situations. 
In the case of semi-conductors such as silicon (Si), germanium (Ge)\cite{Callaway59,Holland63} and gallium arsenide (GaAs)\cite{Waugh63,Bhandari65}, the relaxation times have allowed to compute semi-analytically thermal conductivities in good agreement with measurements. Resolution of the BTE have been achieved on these materials in bulk situations, thin film or superlattice configuration \cite{Majumdar93,Chen97c,Goodson96,Lemonnier00}. At short time scale, these resolutions have been compared to classical solutions\cite{Joshi93,Narumanchi04} and some modifications of the BTE have been proposed \cite{Chen01}. The resolutions based on the Discrete Ordinates Method or on the Finite Volumes Method converge very quickly numerically but have a major drawback : they are governed by a single relaxation time taking into account all the different processes of relaxation such as the anharmonic interactions between phonons, the interactions with impurities and dislocations or the scattering on the material boundaries. The Matthiesen rule which stands that the inverse of the total relaxation term is the sum of the  relaxation times due to every different phenomena is usually used. In the context of the BTE in the relaxation time approximation, this means that all the different interaction or scattering phenomena tends to restore thermal equilibrium. 

An alternative way to solve the BTE is the Monte Carlo method. This method is quite computer time greedy because it necessitates to follow a large number of energy carriers, but it becomes competitive when the complexity of the problem increases, particularly for non-trivial geometries. This method is therefore useful in order to calculate the heat transfer in electronic devices of any shape. Moreover, in this method, different scattering phenomena (impurities scattering, boundary scattering and inelastic scattering) can be treated separately.  The resolution of the BTE by the Monte-Carlo method has been performed for electrons \cite{Jacobini83,Fishetti88,Lugli89,Fishetti93,Pop04,Pop05} but has been little used in the case of phonons. Peterson\cite{Peterson94} performed a Monte Carlo simulation for phonons in the Debye approximation with a single relaxation time. He presented results both in the transient  regime and in equilibrium situation. Mazumder and Majumdar\cite{Mazumder01} followed Peterson's approach but included in their simulation the dispersion and the different acoustic polarization branches. They retrieved both the ballistic and the diffusion situation but did not show any result in the transient regime. Another limit of this last paper is that the $N$ processes and the $U$ processes are not treated separately although they do not contribute in the same way to the conductivity. 

The starting point in our work are these two contributions. We follow individual phonons in a space divided into cells. The phonons after a drift phase are able to interact and to be scattered. The speed and the rate at which phonons scatter depends on the frequency. We ensure that energy is conserved after each scattering process. This procedure is different whether the phonons interact through a $N$ process or a $U$ process. This paper is therefore an improvement of existing phonon Monte Carlo methods and is validated on simple examples such as a semiconductor film heated at two different temperatures. 

Section II recalls the basic hypothesis governing the BTE. Fundamental quantities such as the number of phonons, the energy and the density of states are also defined. The phonons properties are also presented through their dispersion relations. Section III exposes the Monte Carlo method used in this paper. Boundary conditions, phonons drift and scattering  procedures are given in details. Section IV present transient results in the diffusion and ballistic regimes. Thermal conductivities of silicon and germanium between 100 K and 500 K are numerically estimated. Influence of conductivity thermal dependence on heat conduction within a slab is studied. Finally, size effects on phonons transport at very short scales are considered.

\section{Theory}
\subsection{Boltzmann Transport Equation}

The Boltzmann Transport Equation (BTE) is used to model the phonons behavior in a crystal lattice. This equation is related to the variation of the distribution function $f(t,\bf{r},\bf{K})$  which depends on time $t$, location $\bf{r}$ and wave vector $\bf{K}$. $f(t,\bf{r},\bf{K})$ can also be defined as the mean particle number at time $t$ in the $d^3\bf{r}$ volume around $\bf{r}$ with $\bf{K}$ wave vector and $d^3\bf{K}$ accuracy. In the absence of external force, the BTE expression is\cite{Ashcroft76}
\begin{equation}\label{BTE}
\frac{\partial f}{\partial t}+\nabla_{\bf{K}}\omega\cdot\nabla_{\bf{r}}f=\left.\frac{\partial f}{\partial t}\right|_{collision}
\end{equation} 
with the phonon group velocity $\bf{v}_{\rm g}=\nabla_{\bf{K}}\omega$.

Integration of the distribution function over all the wave vectors of the first Brillouin zone and all the locations leads to the phonons number $N(t)$ at a given time in the crystal. The lhs term of Eq.(\ref{BTE}) accounts for the phonons drift in the medium and the rhs term for the equilibrium restoration due to phonons collisions with themselves, impurities and boundaries.

The collisional term modeling is the key point in the BTE resolution. In the case of photons, it can be quite easily modelized by an absorption term, an emission term and an elastic scattering term in which  a scattering phase function relates a photon in the incoming and outgoing propagation directions during a scattering event \cite{Chandrasekhar}. However, in the case of phonons, there is no absorption, nor emission but only scattering events. Scattering events at the borders can simply be treated during the drift phase i.e. when a phonon reaches a border.  Scattering with impurities  can be treated similarly to the isotropic scattering of photons in thermal radiation. Scattering of phonons due to the anharmonic terms of the potential are quite difficult to modelize. We know nevertheless that these terms are responsible of the thermal conductivity i.e. tend to restore thermal equilibrium.  Therefore, in this work we use the Relaxation Time Approximation for three phonons scattering processes. The collision time used in this formalism comes from Normal and Umklapp relaxation times which are further estimated.

\subsection{Lattice modeling}
As it has been exposed previously, the thermal behavior of the crystal can be considered from the phonons characteristics (location, velocity and polarization) within the medium. They might be obtained through the BTE solution since the distribution function can be easily related to the energy and therefore to the temperature. Using an integrated distribution function, one can express the total vibrational energy of the crystal as \cite{Kittel04}
\begin{equation}\label{Eie1}
 	E=\sum_p\sum_{\bf{K}} \left(\left\langle n_{{\bf K},p} \right\rangle+\frac{1}{2} \right) \hbar \omega
\end{equation} 
where $\left\langle n_{{\bf K},p} \right\rangle$ is the local thermodynamic phonons population with polarization $p$ and wave vector $\bf{K}$ described by the Bose-Einstein distribution function
\begin{equation}\label{Bose}
	\left\langle n_{{\bf K}, p} \right\rangle=\frac{1}{\exp{\left(\frac{\hbar \omega}{k_B T} \right)-1 }}
\end{equation}
$E$ is the material volumic energy. It is obtain by summation in Eq.(\ref{Eie1}) of each quantum $\hbar \omega$ over the two polarizations for transverse, longitudinal and optical modes of phonons propagation. Assuming that the phonons wave vectors are sufficiently dense in the $\bf{K}$-space,
the summation over $\bf{K}$ can be replaced by an integral. Moreover, using $D_p(\omega)$ the phonon density of state, we can achieve the integration in the frequency domain. This two modifications yield
\begin{equation}\label{Eie2}
	E=\sum_p\int_{\omega} \left(\left\langle n_{\omega ,p} \right\rangle+\frac{1}{2} \right) \hbar \omega D_p(\omega) g_p d\omega
\end{equation}
with $D_p(\omega)d\omega$ the number of vibrational modes in the frequency range $\left[\omega,\omega+d\omega\right]$ for polarization $p$ and $g_p$ the degeneracy of the considered branch. In the case of a three dimensional crystal ($V=L^3$) we have
\begin{equation}\label{Dens_etat}
	D_p(\omega)d\omega=\frac{d\bf{K}}{\left(2\pi/L \right)^3 }=\frac{V K^2dK}{2\pi^2}
\end{equation}
The $\frac{1}{2}$ term in Eq.(\ref{Eie1}) is the constant zero point energy which do not participate to the energy transfer in the material, therefore it has been suppressed. Using the group velocity definition, Eq.(\ref{Eie2}) might be rewritten
\begin{equation}\label{Eie3}
	E=V\sum_p\int_{\omega} \left[ \frac{\hbar \omega}{\exp{\left(\frac{\hbar \omega}{k_B T} \right)-1 }}\right]  \frac{K^2}{2\pi^2 v_g} g_p d\omega
\end{equation}
The numerical scheme we are going to present is mainly based upon energy considerations. The previous expression Eq.(\ref{Eie3}) will be also used to estimate the material temperature by means of a numerical inversion.

\subsection{Dispersion curves}

\begin{figure}[h]\label{figure01}
	\includegraphics[angle=0,width=3in]{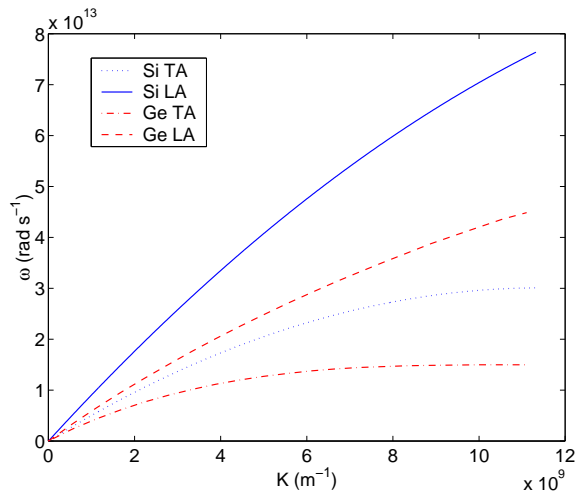}
	\caption{Phonons dispersion curves for silicon and germanium in the first Brillouin zone, ${\bf K}_{max\,Si}=1.1326\cdot10^{10}$ m$^{-1}$ and
	${\bf K}_{max\,Ge}=1.1105\cdot10^{10}$ m$^{-1}$}
\end{figure}

Only a few studies on that topic take into account  dispersion. Indeed, frequency dependence makes calculations longer accounting for velocity variation. However, realistic simulation of phonons propagation through the crystal must take into account interaction between the different branches. Here optical phonons are not considered because of their low group velocity : they do not contribute significantly to the heat transfer. Consequently only transverse and longitudinal branches of silicon and germanium are presented here (Fig. \ref{figure01}). We have made the common isotropic assumption for wave vector and consider the $\left[001\right] $ direction in $\bf{K}$-space. For silicon, we used data obtained from a quadratic fit \cite{Pop04}, whereas germanium experimental curves \cite{Weber77} have been fitted by cubic splines. Phonons group velocity has then been extracted from this data. Note that in silicon an germanium, two acoustic branches have been considered. The transverse branch is degenerate ($g_T=2$) whereas the longitudinal branch is non-degenerate ($g_L=1$).

\section{Monte Carlo method}

The Monte Carlo technique has been widely used in order to solve transport equations. In the heat transfer field, Monte Carlo solutions of RTE are often considered as reference benchmarks. The method accuracy only lies on the number of sample used. Among others, the main advantages of this technique are :
\begin{itemize}
\item The simple treatment of transient problems,
\item The ability to consider complex geometries,
\item The possibility to follow independently each scattering processes (for instance phonon-phonon, phonon-impurity and phonon-boundary processes).
\end{itemize}

The main drawback is computational time. However this method remains a good choice between deterministic approaches such as the Discrete Ordinates Method (DOM) or "exact solution" such as those provided by molecular dynamic which is limited to very small structures.

\subsection{Simulation domain and boundary conditions}

\begin{figure}[h]\label{figure02}
	\includegraphics[angle=0,width=3in]{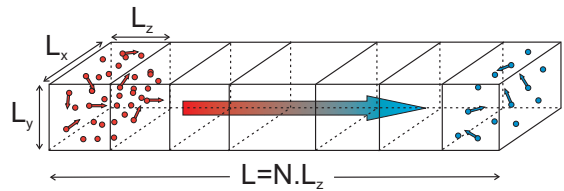}
	\caption{Studied model. Phonons location, energy and velocity are randomly chosen in each cell according to dispersion curves and local temperature.}
\end{figure}

As it was said before, the geometry of the studied material does not matter. Here, a simple cubic cells stack (Fig. \ref{figure02}) is considered since it can be readily related to the plane wall geometry commonly used in thermal problems. Cylindrical cells or multidimensional stacks can be also considered in order to model nanowires or real semi-conductors.

Concerning boundary conditions, we assume that the lateral walls of the cells (in $x$ and $y$ directions) are specularly reflecting in most of the simulation cases. This means that walls are adiabatic and perfectly smooth. Note also that in that case, the dimension in the $x$ and $y$ directions should not change the result in the simulation. Indeed, when reflection is specular on the lateral cells boundaries, the momentum is preserved in the $z$ direction. The heat flux and the temperature along $z$ should thus not be affected.
At both end of the medium, temperature is assumed to be constant. Therefore, energy in the first and the last cell is calculated from equilibrium distribution functions. Incoming phonons in these cells are thermalized at each time step. Consequently theses cells act as blackbodies.

At this stage, an important point is the choice of the three discretizations : temporal, spatial and spectral. Spatial discretization is directly related to the material geometry : usually cells length are about $L_z\sim100\,$ nm for micrometric objects and can be smaller in the case of thin films or nanowires for instance. The time step choice depends on two parameters : the cell size and the group velocity at a given frequency. In order to consider all scattering events and to avoid ballistic jump over several cells, we state that the time step must be lower than $\Delta t<L_z/ V_{g max}$.

The spectral discretization is uniform, we used $N_b=1000$ spectral bins in the range $\left[0, \omega_{LAmax}\right]$. We have checked that larger discretizations do not increase the results accuracy.

\subsection{Initialization}

The first step of the simulation procedure, once medium, geometry and mesh have been chosen, is to initialize the state of phonons within each cell describing the material. Hence, the number of phonons present in each cell is required. It will be obtained considering the local temperature within the cell and using a modified expression of Eq. (\ref{Eie3}). In this equation, energy is given for all the quanta $\hbar\omega$ associated to a spectral bin. Therefore, it can be rewritten to give the total number of phonons in a cell as
\begin{equation}\label{Nph1}
	N=V\sum_{p=TA,LA}\sum_{b=0}^{N_b} \left[ \frac{1}{\exp{\left(\frac{\hbar \omega_{b,p}}{k_B T} \right)-1 }}\right]  \frac{K_{b,p}^2}{2\pi^2 v_{g\,b,p}} g_p\Delta\omega
\end{equation}
The number of phonons obtain with Eq.(\ref{Nph1}) is usually very large, for instance in a $10$ nm silicon cube at $300$ K, $N$ can be estimated around $5.45\cdot10^5$. In the case of nanoscale structures, direct simulations can be achieved if the temperature is relatively low. In the case of microscale samples or multidimensional cell stacking, a weighting factor shall be used to achieved Monte Carlo simulations. Hence, Peterson's \cite{Peterson94} technique has been used. The actual number of phonons $N$ is divided by a constant weight $W$ in order to obtain the number of simulated phonons $N^\star$
\begin{equation}\label{Nph2}
	N^\star=\frac{N}{W}
\end{equation}
In our simulations $W$ maximal values are $W\sim10^4$ for micrometric structures.

During the initialization process, a temperature step is prescribed in the medium. The first cell being raised to the hot temperature $T_h$, the last to the cold one $T_c$. All the phonons in the intermediate boxes are also at $T_c$. Associated theoretical energy in the whole structure is obtained from Eq.(\ref{Eie3}). This energy should match the calculated energy $E^\star$ within all the cells, written into the following form
\begin{equation}\label{Eie4}
	E^\star=\sum_{c=1}^{N_{cell}}\sum_{n=1}^{N^\star}W\times\hbar \omega_{n,c}
\end{equation}
As a consequence, during the initialization, phonons should be added by packs of $W$ at a given frequency, sampled from a normalized number density function $F$. According to Mazumder and Majumdar work \cite{Mazumder01}, this function is constructed doing the cumulative summation of the number of phonons in the $i^{th}$ spectral bin over the total number of phonons Eq.(\ref{Nph1})
\begin{equation}\label{Fcumul}
	F_i\left( T\right)=\frac{\sum_{j=1}^{i}N_j\left( T\right)}{\sum_{j=1}^{N_b}N_j\left( T\right)}
\end{equation}
In this process, a random number $R$ is drawn (all the random numbers discussed here check $0\leq R \leq1$) and the corresponding value $F_i$ gives the frequency $\omega_i$, knowing that  $F_{i-1}\leq R \leq F_i$ location is achieved with bisection algorithm. The actual frequency of the phonon is randomly chosen in the spectral interval prescribing
\begin{equation}\label{freq1}
	\omega_i=\omega_{0,i}+(2R-1)\frac{\Delta\omega}{2}
\end{equation}
where $\omega_{0,i}$ is the central frequency of the $i^{th}$ interval.

Once the frequency is known, the polarization of the phonon has to be determined. It can belong to the TA or LA branch with respect to the Bose-Einstein distribution and the density of states. For a given frequency $\omega_i$, the number of phonon on each branch is: $N_{LA}(\omega_i)=\left\langle n_{LA}(\omega_i) \right\rangle D_{LA}(\omega_i)$ and  $N_{TA}(\omega_i)=2\times \left\langle n_{TA}(\omega_i) \right\rangle D_{TA}(\omega_i)$. The associated probability to find a $LA$ phonon is expressed as
\begin{equation}\label{pola1}
	P_{LA}(\omega_i)=\frac{N_{LA}(\omega_i)}{N_{LA}(\omega_i)+N_{TA}(\omega_i)}
\end{equation}
A new random number $R$ is drawn: if $R<P_{LA}(\omega_i)$ the phonon belongs to the $LA$ branch otherwise it is a transverse one.

The knowledge of the frequency and the polarization leads to the estimation of the phonon group velocity and the phonon wave vector merely using the dispersion curves and their derivatives. Assuming isotropy within the crystal, the direction $\bf{\Omega}$ of the wave vector and the group velocity are obtained from two random numbers $R$ and $R'$ in the Cartesian coordinates $(\bf{i},\,\bf{j},\,\bf{k})$ associated to lengths $L_x$, $L_y$ and $L_z$. Hence $\bf{\Omega}$ is written as 
\begin{equation}\label{direction}
	\bf{\Omega}=\left\lbrace 
	\begin{array}{l}
		\sqrt{1-(2R-1)^2}\cos(2\pi R')\\
		\sqrt{1-(2R-1)^2} \sin(2\pi R')\\
		(2R-1)\\
	\end{array} 
	\right.
\end{equation}

The last operation of the initialization procedure is to give a random position to the phonon within the cell. In the grid previously considered, location of the $n^{th}$ phonon in the cell $c$ is
\begin{equation}\label{position}
	{\bf r_{\rm n,c}}={\bf r_{\rm c}} +  L_x R\, {\bf i} + L_y R'\,{\bf j} + L_z R''\,{\bf k}
\end{equation} 
where $\bf{r_{\rm c}}$ is the coordinates of the cell and $R$, $R'$ and $R''$ three random numbers.

\subsection{Drift}

Once the initialization stage is achieved, phonons are allowed to drift inside the nanostructure. Considering the time step $\Delta t$ and their velocities, each phonon position is updated : $\bf{r_{\rm drift}}=\bf{r_{\rm old}}+\bf{v_{\rm g}}{\rm \Delta t}$. In the case of shifting outside of the lateral boundaries (in $\bf{i}$ and $\bf{j}$ directions) the phonon is specularly reflected at the wall. In the case of diffuse reflection with a particular degree $d$ ($0 \leq d \leq 1$, $d=0$ purely specular, $d=1$ purely diffuse) a random number $R$ is drawn. When $R$ is lower than $d$ a new phonon propagation direction is calculated using Eq.(\ref{direction}).

When a phonon reaches the bottom ($z_{min}$) or the top ($z_{max}$) of a cell, it is allowed to carry on its way in the previous or next cell respectively. As a result, it is going to modify the cell energy and by extension its local temperature. At the end of the drift phase, the actual energy $\widetilde{E^\star}$  is computed in all the cells using Eq.(\ref{Eie4}). Then, the actual temperature $\widetilde{T}$ is obtained with Eq.(\ref{Eie3}) doing a Newton-Raphson inversion \cite{Num_Rec}. Phonons drifting in the first and last cellules are thermalized to the cold or hot temperature in order to keep boundary cells acting as blackbody sources.

\subsection{Scattering}

In the Monte Carlo simulation, the scattering process has been treated independently from the drift. The phonon-phonon scattering aims at restoring local thermal equilibrium in the crystal since it changes phonons frequency. Collisions with impurities or crystal defects as well as boundary scattering do not change frequency but solely the direction $\bf{\Omega}$. These last phenomena are significant when low temperatures are reached and the phonon mean free path becomes large. Here only three-phonon interactions have been considered.

As already said before, there are two kinds of three-phonon processes : Normal processes ($N$) which preserve momentum and Umklapp processes ($U$) which do not preserve momentum by a reciprocal lattice vector. These two mechanisms have consequences on the thermal conductivity of the crystal. When the temperature is sufficiently high ($T\gtrsim T_{Debye}$), $U$ processes become significant and directly modify heat propagation due to the resistivity effect on energy transport. On the other hand, Normal scattering also affects heat transfer since it modifies frequency distribution of the phonons. For phonons described by  $(p,\omega,\bf{K})$ and $(p',\omega',\bf{K'})$ scattering to $(p'',\omega'',\bf{K''})$, the following relations are checked 
\begin{equation}\label{3phonons1}
	\left\lbrace 
	\begin{array}{l}
		\rm{energy\,:}\quad\hbar\omega+\hbar\omega'\leftrightarrow\hbar\omega''\\
		N\, \rm{processes\,:}\quad\bf{K}+\bf{K'}\leftrightarrow\bf{K''}\\
		U\, \rm{processes\,:}\quad\bf{K}+\bf{K'}\leftrightarrow\bf{K''}+\bf{G}\\
	\end{array} 
	\right.
\end{equation}
where $\bf{G}$ is a lattice reciprocal vector. Scattering also involves polarization in the way that acoustic transverse and longitudinal phonons can interact. According to Srivastava \cite{Srivastava90} for $N$ and $U$ processes different combinations are possible
\begin{equation}\label{3phonons2}
	\left\lbrace 
	\begin{array}{l}
		N\,{\rm and}\,U\, \rm{processes\,:}\quad T+T \rightleftharpoons L\quad and \quad T+L \rightleftharpoons L\\
		N\, \rm{processes\,only\,:}\quad T+T \rightleftharpoons T\quad and \quad L+L \rightleftharpoons L\\
	\end{array} 
	\right.
\end{equation}
For the $N$ processes only, all the participating phonons must be collinear to achieve scattering. Usually these interactions are neglected.

Direct simulation of phonons scattering is an awkward challenge. With Monte Carlo simulations, it is possible to modelize phonons collisions with neighbors  as in the gas kinetic theory calculating a three particles interaction cross section. However, in the present study, the frequency discretization might not be sufficiently thin to assess every three phonons processes. Thus the collisional process is treated in the Relaxation Time Approximation. Several studies on that topic have been carried out since the early work of Klemens \cite{Klemens51}, a detailed paper of Han and Klemens \cite{Han93} recalled them.

Relaxation times $\tau$ have been proposed for several crystals. They depend on the scattering processes, the temperature and the frequency. Holland's work on silicon \cite{Holland63} and the recent study of Singh for germanium \cite{Singh03} provide various $\tau$ values. The independence of the scattering processes is used to consider a global three phonons inverse relaxation times accounting for $N$ and $U$ processes $\tau_{NU}$. It has been obtained using the Mathiessen rule ($\tau_{NU}^{-1}=\tau_{N}^{-1}+\tau_{U}^{-1}$).

In order to be implemented in the Monte Carlo simulation, the scattering routine requires an associated collision probability $P_{scat}$. This one is derived saying that the probability for a phonon to be scattered between $t$ and $t+dt$ is $dt/\tau$. Thus,
\begin{equation}\label{Pcol}
	P_{scat}=1-\exp\left(\frac{-\Delta t}{\tau_{NU}}\right) 
\end{equation} 
A random number $R$ is drawn, if $R<P_{scat}$ the phonon is scattered. As a result new : frequency, polarization, wave vector, group velocity and direction have to be resample with respect of energy and momentum conservation. 

In previous studies on that topic,\cite{Peterson94, Mazumder01} the frequency sampling after collision was achieved from the normalized number density function $F$ at the actual temperature $\widetilde{T}$ of the cell obtained at the end of the drift procedure. In this approach the actual energy after the scattering stage is usually different from the "target" one obtained with temperature $\widetilde{T}$. Hence a subsequent "creation/destruction" scheme is necessary to ensure energy conservation. In  fact, in the preceding procedure at thermal equilibrium, the probability of destroying a phonon of frequency $\omega$ and polarization $p$ is different from the probability of creating this phonon. This means that the Kirchhoff law (creation balances destruction) is not respected. In order to create phonons at the same rate they are destroyed at thermal equilibrium, the distribution function used to sample the frequencies of the phonons after scattering has to be modulated by the probability of scatterring. So we define a new distribution function
\begin{equation}\label{Fcumulscat}
	F_{scat}(\widetilde{T})=\frac{\sum_{j=1}^{i}N_j\left( \widetilde{T}\right)\times P_{scat\, j}}{\sum_{j=1}^{N_b}N_j\left( \widetilde{T}\right)\times P_{scat\, j}}
\end{equation}

\begin{figure}[h]\label{figure03}
	\includegraphics[angle=0,width=3in]{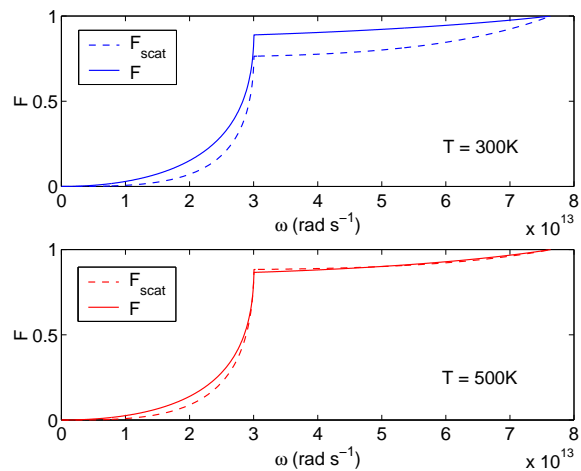}
	\caption{Normalized number density function in silicon with and without $P_{scat}$ correction}
\end{figure}

Taking into account the scattering probability in the distribution function $F_{scat}$ ensures that a destructed phonon on both transverse and longitudinal branches can be resample with a not to weak energy as it can be seen (Fig. \ref{figure03}).

\begin{figure}[h]\label{figure04}
	\includegraphics[angle=0,width=3in]{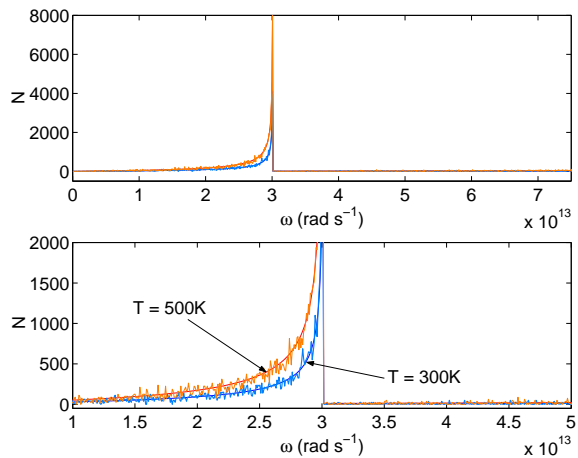}
	\caption{Frequency spectra at 300K and 500K for silicon}
\end{figure}

According to the described simulation procedure after initialization step, phonons in cell $c$ are described by [$T_c$, $F(T_c)$, $N^{\star}(T_c)$, $E^{\star}(T_c)$]. They are allowed to drift and the state of cell $c$ before scattering is [$\widetilde{T_c}$, $F(T_c)$, $N^{' \star}(\widetilde{T_c})$, $E^{' \star}(\widetilde{T_c})$]. Then three-phonons collisions occur and change energy by frequency resetting of the colliding phonons (using the distribution function $F_{scat}$, leading to the final state [$\widetilde{T_c}$, $F_{scat}(\widetilde{T_c})$, $N^{' \star}(\widetilde{T_c})$, $E^{'' \star}(\widetilde{T_c})$]. Hence energy can be express as :
\begin{equation}\label{Eie5}
	E^{'' \star}=\underbrace{\sum_{i=1}^{N^{' \star}_{scat}} \hbar\widetilde{\omega_i}}_{\Rightarrow F_{scat}(\widetilde{T_c})}+\underbrace{\sum_{i=1}^{N^{' \star}-N^{' \star}_{scat}} \hbar\omega_i}_{\Rightarrow F(T_c)}
\end{equation} 
Furthermore the number of colliding phonons can simply be express as $N^{' \star}_{scat}(\widetilde{T_c})=P_{scat}\times N^{' \star}(\widetilde{T_c})$. One then sees, that, if we want to preserve energy at thermal equilibrium, the new normalized number density function $F_{scat}$ must take into account this collisional probability. Energy conservation during Monte Carlo simulation might be noticed from frequency distribution (Fig. \ref{figure04} that matches the theoretical distribution give by Eq.(\ref{Nph1}).
Concerning momentum conservation, the task is harder to address since the Monte Carlo process considers phonons one by one. Consequently triadic $N$ or $U$ interactions cannot be rigorously treated. In a first approach, we propose the following procedure to take into account the fact that $U$ processes contribute to the thermal resistance whereas the $N$ processes do not. When the phonons scatter through a $U$ process, their directions after scattering are randomly chosen as in the initialization procedure. Therefore, these phonons are randomly scattered and contribute to the diffusion of heat. On the contrary, it is assume that scattering phonons experiencing a $N$ process do not change their propagation direction $\bf{\Omega}$. 

Statistically, for a given temperature and frequency, the phonons are destroyed by scattering at the same rate they appear. A phonon which scatters has a great chance to be replaced in the computation by a phonon of a near frequency. Therefore, by this treatment, the $N$ processes "approximately" preserve momentum. Nevertheless, a more accurate treatment should be done in order to respect exactly the momentum in the $N$ processes. For a plan-parallel geometry, it seems possible to guarantee the momentum conservation in a single direction.

In fact, the relaxation time estimation \cite{Han93} states that there is a frequency limit $\omega_{limit}$ for the transverse acoustic branch. $\omega_{limit}$ actually corresponds to ${\bf K}={\bf K}_{max}/2$. Below this limit frequency, there is no $U$ processes. On the other hand for $\omega > \omega_{limit}$, $N$ processes are no longer considered and the propagation direction must be resampled in the case of a collision. In what concerns the longitudinal acoustic branch, there is no limit frequency. According to Holland \cite{Holland63} only $N$ processes exist. However applying this assumption implies that momentum has to be conserved for each scattering event involving a $LA$ phonon. This leads to thermal conductivity values higher than the theoretical ones for temperatures between 100 K and 250 K. In order to ensure a more realistic momentum conservation we set that half of the colliding phonons keep their original $\bf{\Omega}$, the others ($U$ processes) are directionally resampled.

\section{Results and discussion}

Different kinds of simulations have been performed so as to check the computational method. Tests in both diffusion and ballistic regimes are carried out for silicon and germanium. Moreover, if small thermal gradients are considered, one can estimate the thermal conductivity $k$ from the heat flux through the structure. This has been realized for Si and Ge between 100 K and 500 K.

Knowing that the conductivity varies with temperature according to a power law in the case of Si and Ge for $T$ greater than 100 K, it is obvious that a large thermal gradient applied to our media should not bring a purely linear solution. Hence simulations in this specific case have been done. The model ability to correctly predict steady state has been confirmed with comparison to analytical solution.

Eventually, we studied size effects on thermal behavior of nanostructures. It appears that the ballistic regime can be retrieved at room temperature when the sample size is close to the nanometer scale.

\subsection{High temperature transient calculations}

\begin{figure}[h]\label{figure05}
	\includegraphics[angle=0,width=3in]{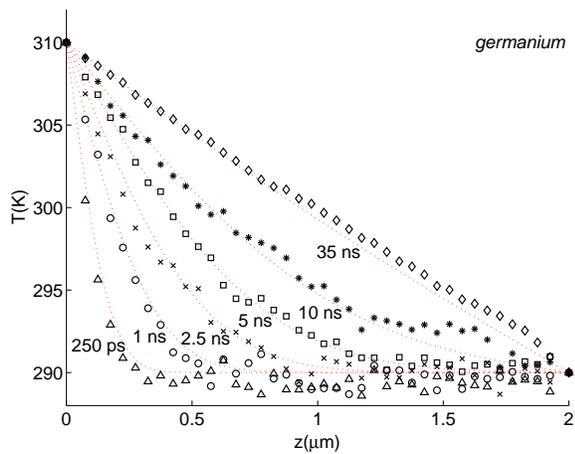}
	\caption{Transient temperature in Fourier's regime for germanium and comparison with the analytical solution of heat conduction equation with a constant thermal diffusivity ($\alpha_{Ge}=0.36\cdot10^{-4}\,{\rm m^2s^{-1}}$) (dotted curves)}
\end{figure}

Concerning high temperature transient calculations, the simulated case is described by the following parameters :
\begin{itemize}
\item Hot and cold temperatures: $T_h=310$ K and $T_c=290$ K,
\item Medium geometry: stack of 40 cellules ($L_x=L_y=5\cdot10^{-7}$ m, $L_z=5\cdot10^{-8}$ m),
\item Time step and spectral discretization: $\Delta t=5$ ps and $N_b=1000$ bins,
\item Weighting factor: $W=3.5\cdot10^4$ for Si and $W=8\cdot10^4$ for Ge.
\end{itemize}
Both materials were tested. Germanium calculation results are presented here (Fig. \ref{figure05}). In order to assess the Monte Carlo solution, transient theoretical comparison exists in the case of the Fourier limit. Nevertheless, it requires that the thermal diffusivity $\alpha$ remains constant. In the chosen temperature range, according to the IOFFE database \cite{ioffe}, Ge thermal diffusivity is equal to $\alpha=0.36\cdot10^{-4}\,{\rm m^2s^{-1}}$.

The considered test case has been described in \"Ozi\c{s}ik's book \cite{Ozisik_bk1} on heat conduction equation. Within the described structure heat transfer is along $z$ axis and analytical solution for one-dimensional medium could be obtained from integral transform. Temperature distribution in the slab is given by an infinite sum that requires enough terms in the case of short time calculation. However a simpler analytical solution might be obtained with Laplace's transform
\begin{eqnarray}\label{Tslab}
	\frac{T(z,t)-T(L,t)}{T(0,t)-T(L,t)}&=&\left[{\rm erfc}\left(\frac{z}{2\sqrt{\alpha t}} \right) -{\rm erfc}\left(\frac{2L-z}{2\sqrt{\alpha t}} \right)\right. \nonumber\\
	&+&\left. {\rm erfc}\left(\frac{2L+z}{2\sqrt{\alpha t}} \right)\right] 
\end{eqnarray} 
with erfc the complementary error function. The theoretical solution is only valid for short time and its accuracy is better than $1\%$ if the Fourier number ($Fo=\alpha t/L^2$) check $Fo\leq0.7$. In the case of a $2\,\mu$m germanium slab it leads to $t\leq 78$ ns, which is largely enough to reach steady state. 

The calculated values have been obtained from ten simulations averaged Fig. \ref{figure05}, the random numbers seed being reset for each computations. Monte Carlo model ability to predict correctly temperature from the first moments till steady state is clearly illustrated. The remaining noise can be reduced with lower values of $W$ weighting factor. Diffusion regime is obtained after $30$ ns. Similar results are obtained for silicon, however the modelized slab has to be larger ($L=4\mu$m) because ballistic effects are observed near the cold limit. This point will be discussed latter.

\subsection{Low temperature transient calculations}

For low temperatures, heat transport inside the slab is different since phonons interactions change. In this case $U$ collisions are neglectibles and the only resistive processes are assigned to impurities, defects and boundary scattering. These phenomena have to be carefully examined in the case of thermal conductivity estimation below 100 K. 
In fact, for very low temperatures the phonon mean free path grows and becomes larger than the structure length. Hence, phonons can travel from hot to cold extremity without colliding. This is the ballistic regime similar to the one observed with photons exchanged between two black plates at different temperatures \cite{Heaslet65}. In this peculiar case temperature in steady state is equal to the following constant value
\begin{equation}
	T_{ballistic}=\left[\frac{T_h^4+T_c^4}{2} \right]^{1/4}
\end{equation} 
The simulation case parameters are
\begin{itemize}
\item Hot and cold temperatures: $T_h=11.88$ K, $T_c=3$ K and $T_{ballistic}=10$ K,
\item Medium geometry: stack of 40 cellules ($L_x=L_y=5\cdot10^{-7}$ m, $L_z=2.5\cdot10^{-7}$ m),
\item Time step and spectral discretization: $\Delta t=5$ ps and $N_b=1000$ bins,
\item Weighting factor: $W=20$ for Si and $W=30$ for Ge. 
\end{itemize}

\begin{figure}[h]\label{figure06}
	\includegraphics[angle=0,width=3in]{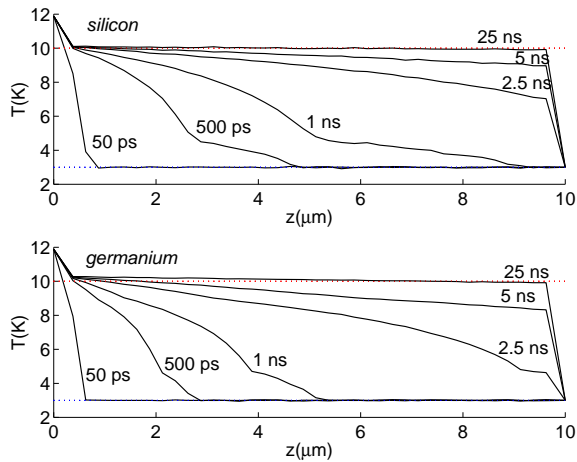}
	\caption{Transient temperature in the ballistic regime for silicon and germanium}
\end{figure}

Results for silicon and germanium (Fig.\ref{figure06}) give the expected results for the ballistic limit. It can be noticed that the current representation exhibits an artificial link between black boundaries and the first medium cell due to the spatial discretization.
It can be seen that hot phonons do not fly straight toward the cold limit. More than 1 ns is necessary to heat the last cell in the case of silicon. This is in agreement with velocities prescribed by dispersion curves. Heat propagation in germanium is slower since phonons group speed is also lower. Results at low temperatures obtained with our method have been already be predicted by Joshi and Majumdar \cite{Joshi93} in similar cases, who applied successfully the Equation of Phonon Radiative Transfer (EPRT) in ballistic regime.

\subsection{Si and Ge thermal conductivities}

There are several ways to perform the thermal conductivity calculation of a semiconductor. Among these techniques, Holland's method \cite{Holland63} based on phonons kinetic theory was largely employed. Molecular dynamic simulations can be also used to obtain $k$. In the present study thermal conductivity has been determined knowing the heat flux (phonon energy transport) through the medium for a given thermal gradient directly applying Fourier's law. As in Mazumder's work \cite{Mazumder01}, the temperature difference between hot ans cold extremities is set to 20 K so as to determine average conductivities. The phonon heat flux is calculated along $z$ axis according to the following relation
\begin{equation}\label{flux_z}
	\phi=\sum_{n=1}^{N^\star}W\hbar \omega_{n}\bf{v_{\rm g\,n}}\cdot \bf{k}
\end{equation} 

\begin{figure}[h]\label{figure07}
	\includegraphics[angle=0,width=3in]{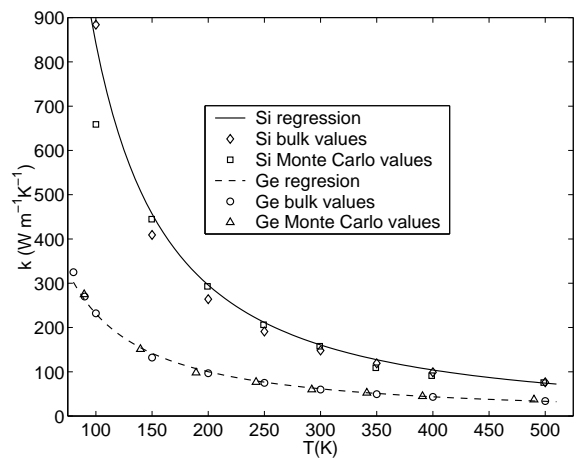}
	\caption{Silicon and germanium thermal conductivities; comparison between bulk theoretical values and Monte Carlo calculated values.}
\end{figure}

Simulations have been carried on between 100 K and 500 K (Fig.\ref{figure07}) on 2 $\mu$m thick samples.

Comparison of the Monte Carlo calculated conductivities is achieved with bulk data. Solid and dash curves are linear power law regression of theoretical data in the considered thermal range. These values are used in next analytical calculations. We have appraised for 200 K $\leq T \leq$ 600 K
\begin{eqnarray}\label{conduc_Si_Ge}
&&\lambda_{Si}\left( T\right)=\frac{\exp\left(12.570 \right) }{T^{1.326}}\nonumber \\
&&\lambda_{Ge}\left( T\right)=\frac{\exp\left(10.659 \right) }{T^{1.150}}
\end{eqnarray} 

For germanium a very good agreement is obtained with bulk values in the whole temperature domain. The maximum relative error being under $8\%$. In this case, at 100 K, phonon mean free path is lower to the micrometer according to Dames \cite{Dames04}. Consequently the stucture size is large enough to assume the acoustic thick limit. Furthermore, this calcul benefits from recent relaxation time estimation which have been fixed with a good precision \cite{Singh03}. The influence of these factors on calculated conductivity is usually strong. Silicon results are also close to the bulk ones till 150 K where the relative error is equal to $7\%$. For lower temperatures, discrepancy between theory and simulation increases. This gap can be assigned to size effects since phonon mean free path grows when temperature is falling. Here, it becomes similar to the slab size. Yet, if we refer to Asheghi \cite{Asheghi02} work on thin films thermal conductivity, at temperatures below 100 K, significantly decreases in comparison with bulk due to stronger reduction of phonon mean free path by boundaries. Actually for pure 3 $\mu$m silicon film thermal conductivity is close to $600\,{\rm W m^{-1}K^{-1}}$ at 100 K \cite{Asheghi02}. This value is comparable to the $658\,{\rm W m^{-1}K^{-1}}$ obtained for our 2 $\mu$m film by Monte Carlo simulation.

\subsection{Effect of non linear conductivity}

In this fourth part, transient simulations with samples heated under a large thermal step have been conducted. The purpose of such calculations was to underling the model capacity to correctly predict steady state when medium properties ($k(T)$) vary with temperature. In the previous part thermal conductivities of both bulk materials have been estimated with a power law Eq.(\ref{conduc_Si_Ge}). Hence, the analytic solution for temperature profile within a slab can be easily determined in steady state by the resolution of a first order differential equation as
\begin{equation}\label{T_stat}
	T(z)_{ Steady\, state}=\left[\left( \frac{z}{L}\right)T_c^ {\left(\gamma+1\right)} +\left( 1-\frac{z}{L}\right)T_h^ {\left(\gamma+1\right)}   \right] ^{1/(\gamma+1)}
\end{equation} 
where conductivity can be written as $\lambda(T)=C\times T^\gamma$.
 
\begin{figure}[h]\label{figure08}
	\includegraphics[angle=0,width=3in]{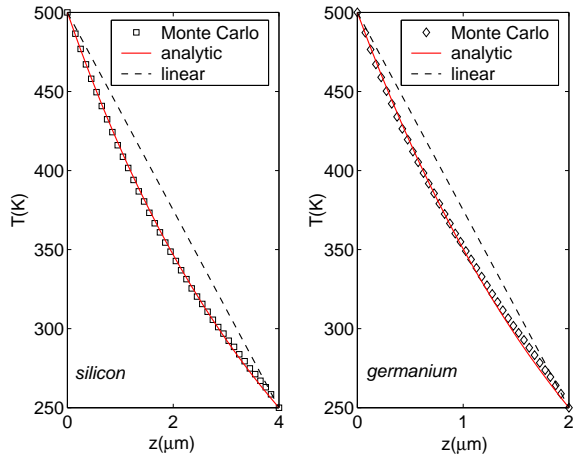}
	\caption{Steady state temperature in Fourier's regime for silicon and germanium in the case of a large thermal gradient; comparison to the heat conduction equation analytical solution for temperature dependent conductivity.}
\end{figure}

In order to avoid boundary effects in the case of silicon, $4\,\mu$m thick sample, with larger cells ($L_z=1\cdot 10^{-7}$m), is used. The initial geometry is kept for germanium slab. Temperatures are now : $T_h=500$ K and  $T_c=250$ K. Time step remains equal to $5$ ps. In both cases (Fig.\ref{figure08}) Monte Carlo simulations give a very good estimation of the steady state behavior. Results are averaged over five computations on the last 1000 time steps (i.e. after 45 ns of elapsed time). At the cold limit of germanium sample a weak deviation exits between simulation and theory. The relative error on temperature remains smaller than $1.5\%$, in this area. This mismatch could be assigned to boundary effects, associating diffusive and ballistic regimes near the limits as we will detailled it in the last part. It could also be due to the accuracy of the bulk thermal conductivity fitting at low temperatures.

Besides, inversion of such curves can theoretically provide the variation of $k$ on a given thermal range, as long as the medium is in the acoustic thick limit.

\subsection{Size effect on heat diffusion}

From the previous calculations, it is obvious that the phonon mean free path modification with the temperature acts as a major factor in heat conduction. So, if the structure size is adjusted in order to match the mean free path at any temperature, ballistic phenomena should be observed. In this study only silicon is used and the simulation parameters are 
\begin{itemize}
\item Hot and cold temperatures: $T_h=310$ K and $T_c=290$ K,
\item Number of cells: 40,
\item Total length and time step ($L,\,\Delta t$): (2 nm ,$5\cdot10^{-3}$ ps), (20 nm ,$5\cdot10^{-2}$ ps), (200 nm ,$5\cdot10^{-1}$ ps), (2$\mu$m ,5 ps) and (4$\mu$m ,5 ps).
\end{itemize}

Weighting parameters and lateral cells lengths are adjusted in order to keep approximatetively 18000 phonons in each cell.

\begin{figure}[h]\label{figure09}
	\includegraphics[angle=0,width=3in]{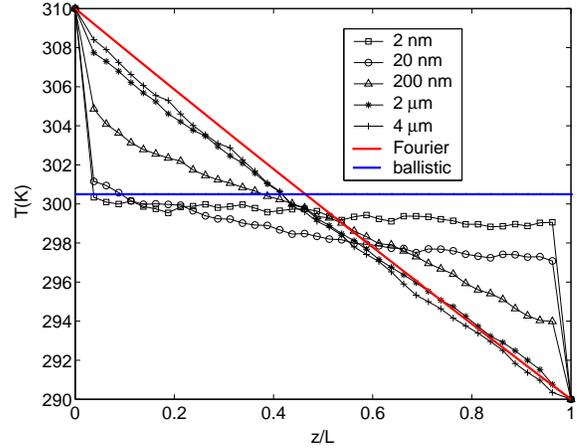}
	\caption{Steady state temperature for silicon, influence of the slab thickness; comparison to the analytical solution in the diffusive and ballistic limits.}
\end{figure}

Temperature profiles when steady state is reached have been plotted for each sample versus adimensionnal length $z/L$. Comparison to diffusive and ballistic regime is displayed in Fig.\ref{figure09}. With the imposed boundary temperatures the ballistic limit is equal to $T_{ballistic}=300.5$ K.
In the case of structures length lower than 200 nm ballistic trend mixed to phonons diffuse transport is observed. The temperature profile gets closer to the ballistic limit for sample size around the nanometer scale. Nevertheless, this approximately represents ten atom layers and therefore might encounter the modelization limitation. On the contrary, in a silicon sample thicker than 4 $\mu$m, temperature reaches the Fourier's regime and can be similarly obtained with heat conduction equation at least cost.

\section{Conclusions}

An improved Monte Carlo scheme that allows transient heat transfer calculations at time and space nanoscales, on the basis of phonons transport, has been presented. This model accounts for phonons transitions between longitudinal and transverse acoustic branches and can be simply applied to several semiconductors if their dispersion relations are known. A particular attention has been paid to the energy and momentum conservation during collision process.

Numerical results forecast have been assessed in different heat transfer modes. In slab configuration, a good agreement was found for both extreme phonon motion which are the diffusive and ballistic ones. Bulk thermal conductivities of silicon and germanium have been numerically retrieved with a maximal error lower than $8\%$. Besides, our Monte Carlo model correctly predict temperature profile in more peculiar situations, when strong thermal gradient or very small sizes are encountered.

Nevertheless some key points need to be refined. Among them momentum conservation procedure might be improved, especially for one dimensional applications. Optical phonons effect on heat transport were neglected. However according to the recent study of Narumanchi \cite{Narumanchi04} they must be taken into account especially for capacitive properties prediction. Regarding the collision process, improvements might be expected. Using theoretical values of $\tau$ recalled by Han and Klemens \cite{Han93}, direct calculation of phonon scattering relaxation time can be realized in each authorized spectral bin. Hence, a more realistic approach of three phonons interactions should be achieved.

We are currently working on this improvements but also on other potential implementation of the method such as those related to the superlattices and the nanowires.

\section*{Acknowledgments}

The authors would like to thank Dr. Denis Lemonnier (LET - ENSMA) for critical comments and helpful discussions.

\bibliography{phonons080405}

\end{document}